\documentclass[10pt,a4paper,reqno]{amsart}
\usepackage[margin=2cm]{geometry}

\usepackage{amsmath,amssymb,multirow,graphicx}

\title{A Nonparametric and Functional Wombling Methodology}
\author{Luke A. Barratt, John A. D. Aston}
\date{}

\DeclareMathOperator{\tr}{tr}
\DeclareMathOperator{\sign}{sign}

\renewcommand{\epsilon}{\varepsilon}
\renewcommand{\theta}{\vartheta}
\renewcommand{\phi}{\varphi}

\begin{document}

\begin{abstract}
Wombling methods, first introduced in 1951, have been widely applied to detect boundaries and variations across spatial domains, particularly in biological, public health and meteorological studies. Traditional applications focus on finite-dimensional observations, where significant changes in measurable traits indicate structural boundaries. In this work, wombling methodologies are extended to functional data, enabling the identification of spatial variation in infinite-dimensional settings. Proposed is a nonparametric approach that accommodates functional observations without imposing strict distributional assumptions. This methodology successfully captures complex spatial structures and discontinuities, demonstrating superior sensitivity and robustness compared to existing finite-dimensional techniques. This methodology is then applied to analyse regional epidemiological disparities between London and the rest of the UK, identifying key spatial boundaries in the shape of the first trajectory of Covid-19 incidence in 2020. Through extensive simulations and empirical validation, demonstrated is the method's effectiveness in uncovering meaningful spatial variations, with potential applications in a wide variety of fields.
\end{abstract}

\maketitle

\section{Introduction}

In 1951, Womble considered the differentiation of populations by examining the variation of traits across geography. \cite{womble1951differential} In particular, a measure of total variation of $p$ traits $(g_j)_{j\in[p]}$ across a transect of space $s$ is provided as $\sum_{j=1}^p w_j\left|{\mathrm{d}g_j}/{\mathrm{d}s}\right|$. Such quantities have become known as wombling measures, and large wombling measures are considered indicative of a boundary between geographies of differing structure. Although the original application was in biology---differentiating population groups---similar approaches have been used in various other disciplines such as meteorology, where such analysis elucidates for example the effect of mountain ranges on differentiating weather phenomena on either side. However, such applications have so far been limited to at most finitely dimensional observations; what if it is the variation of functional observations across space that is of relevance?

Of particular interest here is an application to Covid-19 data, namely the phase variation of the first wave in the United Kingdom. The phase variation encodes the aspects of the shape of the first Covid-19 wave beyond the pure amplitude, eg earliness vs lateness, sharpness vs flatness. The phase variation has been estimated in a spatially aware way already by Barratt and Aston. \cite{barratt2024exploring} To summarise, for each of 380 local authorities in the UK, there is an increasing bijection of $[0,1]$ with identity mean which links calendar time to the underlying progression through a typical wave shape. In their paper, exploratory analysis suggests that the London region of England exhibits substantially different behaviour than the rest of the country, and this is backed up by one-dimensional summaries of these phase variation functions, which demonstrate quantitatively that London in general experienced an earlier and sharper wave of Covid-19. Sought is a way to analyse whether the boundary around London is indeed significantly larger than might be expected at random, taking into account the full functional information available.

This is attempted in two ways: first by developing an existing inferential procedure of Banerjee and Gelfand that works well in a univariate Bayesian framework. \cite{banerjee2006bayesian} It is extended to a multivariate Bayesian framework for the loadings of an appropriate truncated functional basis expansion of the observed data. This approach has various drawbacks in assumption, approximation and computational cost---as well as not being a fully functional methodology---which inspires the development of a fully functional nonparametric methodology. To compare the efficacy of the two methodologies at identifying true from false boundaries, simulations are run and demonstrate that the nonparametric and functional methodology for wombling outperforms the Bayesian one, as well as imposing a fraction of the computational cost. When this methodology is finally applied to the Covid-19 data, it is clear that the boundary around London is indeed significant relative to other regional boundaries in the UK, which may for example justify separate modelling of Covid-19 dynamics inside and outside of London.

\section{Bayesian Wombling}

First will be briefly summarised the approach to univariate gradient estimation and wombling of Banerjee and Gelfand, before this approach is extended to the functional setting. \cite{banerjee2006bayesian} This will provide a reference standard approach to which the novel nonparametric approach will be compared.

\subsection{Univariate framework and methodology}

Suppose $Z$ is a $\mathbb{R}$-valued mean square differentiable spatial process on $\mathcal{S}\subset\mathbb{R}^2$, i.e. $Z:\mathcal{S}\rightarrow\mathbb{R}$. (Dependence on the $\omega\in\Omega$, the event space, will be suppressed throughout for clarity.) Mean square differentiability implies the existence of a spatial derivative process $\nabla Z:\mathcal{S}\rightarrow\mathbb{R}^2$, satisfying:
\begin{equation}
\forall s\in\mathcal{S},u\in S^1,\quad \lim_{\delta\rightarrow 0} \mathbb{E}\left(\frac{Z(s+\delta u) - Z(s)}{\delta} - u^\top\nabla Z(s)\right)^2 = 0,
\end{equation}
where $S^1$ is the one-sphere according to the $\ell^2$ norm. Ultimately, of interest is the average rate of change of the spatial process $Z$ across a finite, one-dimensional boundary $\mathcal{C}$. This can be understood as the normalised flux of $\nabla Z$ across the boundary, which is defined as the normalised wombling measure of $Z$ at $\mathcal{C}$:
\begin{equation}
\bar{W}_Z(\mathcal{C}) := \frac{1}{|\mathcal{C}|}\int_\mathcal{C} n_\mathcal{C}^\top(s) \nabla Z(s)\mathrm{d}\ell(s),
\end{equation}
where $|\mathcal{C}|=\int_\mathcal{C}\mathrm{d}\ell$ is the arc length of $\mathcal{C}$, $n_\mathcal{C}^\top(s)$ is the unit normal vector to the curve $\mathcal{C}$ at $s$, and $\ell$ is the arc length measure. The work of Banerjee and Gelfand demonstrates how this normalised wombling measure can be inferred from the data under a Bayesian framework.

To do this, they assume the observable data $Z$ comes from a Gaussian model:
\begin{equation}
Z\sim \mathrm{Gaussian}(\mu,\sigma),
\end{equation}
where $\mu:=\mathbb{E}Z$ is the mean function and $\sigma:=\mathbb{E}\left[(Z-\mu)\otimes(Z-\mu)\right]$ is the variance function, and where $\otimes$ is the outer product on $L^2(\mathcal{S})$. While the inference may be carried out in the general setting, for notational convenience here, and as utilised in the application, it will be assumed that $Z$ is second-order stationary. This means that $\mu(s)=\mu$ is a constant, and $\sigma(s,s')=\sigma(s-s')$ is a function of the displacement between the two locations alone. Excusing the abuse in notation, henceforth only the latter understanding of $\mu$ and $\sigma$ will be used. It will be supposed that observations of $Z$ are made at the locations $(s_i)_{i\in[n]}$, and let $Z_i:=Z(s_i)$.

It can be shown that the posterior of $\bar{W}_Z(\mathcal{C})$ conditional on the $(Z_i)_{i\in[n]}$ is itself is Gaussian:
\begin{multline}
\bar{W}_Z(\mathcal{C})\ |\ (Z_i)_{i\in[n]}\sim \mathrm{Gaussian}\left( \frac{1}{|\mathcal{C}|}\int_\mathcal{C} n_\mathcal{C}^\top(s) G^\top(s)\Sigma^{-1}\begin{pmatrix}Z_1 - \mu \\ \vdots \\ Z_n - \mu\end{pmatrix}\mathrm{d}\ell(s),\right.\\
\left.-\frac{1}{|\mathcal{C}|^2}\int_\mathcal{C}\int_\mathcal{C} n_\mathcal{C}^\top(s) \left(\mathrm{H}\sigma(s-s') + G^\top(s)\Sigma^{-1}G(s')\right) n_\mathcal{C}(s')\mathrm{d}\ell(s)\mathrm{d}\ell(s')\right),
\end{multline}
where $G_{ij}(s) = \nabla_j\sigma(s-s_i)$ for $i\in[n]$, $j\in[2]$ and $s\in\mathcal{S}$, and $\Sigma_{ii'}=\sigma(s_i-s_{i'})$ for $i,i'\in[n]$. To infer on multiple such Wombling measures, one has that they're jointly Gaussian with cross-covariance given by:
\begin{equation}
\mathrm{Cov}\left(\bar{W}_Z(\mathcal{C}),\bar{W}_Z(\mathcal{C}')\ |\ (Z_i)_{i\in[n]}\right) =\\ 
-\frac{1}{|\mathcal{C}||\mathcal{C}'|}\int_{\mathcal{C}'}\int_\mathcal{C} n_\mathcal{C}^\top(s) \left(\mathrm{H}\sigma(s-s')+G^\top(s)\Sigma^{-1}G(s')\right)n_\mathcal{C'}(s')\mathrm{d}\ell(s)\mathrm{d}\ell(s').
\end{equation}

Thus, if the covariance structure is a function of some parameter vector $\theta$ (and if the mean is unknown there is the mean $\mu$ random too), in the Bayesian framework, one can proceed as follows: sample from the posterior distribution $(\theta,\mu)\ |\ (Z_i)_{i\in[n]}$, such as using Monte Carlo techniques like the Metropolis--Hastings algorithm; then, for each obtained sample, plug the parameters with the observed $(Z_i)_{i\in[n]}$ into the above formulae for the mean and covariance of the distribution of $\bar{W}_Z(\mathcal{C})\ |\ \left((Z_i)_{i\in[n]},\theta,\mu\right)$ (or for all curves on which the wombling measure is to be inferred, with their cross-covariances); finally, sample from this distribution for an overall sampling from the posterior distribution $\bar{W}_Z(\mathcal{C})\ |\ (Z_i)_{i\in[n]}$ (or the joint posterior of the wombling measures of the multiple curves).

It can then be decided which curves may be considered wombling boundaries by, for example, seeing whether zero is contained in a size-$(1-\alpha)$ credible interval for the wombling measure; alternatively, standard scores can be computed by dividing the posterior mean by the posterior standard deviation for each curve, and these quantities may be compared.

The above methodology can be quite simply extended, either by including a regression term in the mean structure or by adding a white noise nugget process (which must be jointly Gaussian with the mean square differentiable process). The resulting calculations remain largely analogous.

\subsection{Functional framework and methodologies}

If instead of a univariate spatial process, $Z$ represents a (Gaussian) functional spatial process ($Z:\mathcal{S}\longrightarrow L^2(\mathcal{T})$ for some functional domain $\mathcal{T}$), similar derivations to the above can be made. However, to ensure the model is identifiable and parsimonious (required due to the great computational cost due the Monte Carlo step in posterior estimation, as well as in the subsequent linear algebra and quadrature), simplifying assumptions are required. First, a finite-dimensional representation of the observed functions must be established so that likelihood-based methods can be applied. A mathematically convenient choice is a truncated orthonormal basis expansion, and for the sake of model parsimony it will be useful to have a basis expansion whereby the loadings are independent, i.e. the (functional) principal components. These can be estimated from the data and used directly in the model as an approximation; moreover, a further simplifying approximation could be made since the variances of these loadings can be readily estimated too, improving the parsimony of the Bayesian setup further. However, it is nontrivial taking into account the uncertainty contributed by the functional principal component estimation step, and due to the estimation the assumption that the loadings are uncorrelated is almost surely false. Alternatively, one might extrinsically justify a particular basis expansion. For example, if the observations are stationary in the functional domain, the Fourier expansion has independent loadings. For the derivatives of warping function estimates in the application supplying $Z$, it is known that $\int_\mathcal{T} Z(s;t)\mathrm{d}t=1$ for all $s\in\mathcal{S}$, which fixes the constant Fourier component, making it a particularly attractive expansion in this case. In either case, some simplifying assumptions or approximations must be made for tractability and parsimony.

Suppose the obtained basis expansion is of the form $Z_i \approx \sum_{j=1}^p \alpha_{ij}\phi_j$, whose $(\alpha_{ij})_{j\in[p]}$ are realisations of a multivariate spatial process $\alpha:\mathcal{S}\rightarrow \mathbb{R}^p$ at locations $s_i$ for $i\in[n]$; there will be notated the component spatial processes $\alpha_j(s):=(\alpha(s))_j$, which must be mean square differentiable. As described above, there will be the assumption that $\alpha_{ij}\perp \alpha_{i'j'}$ for $j\neq j'$ (indeed uncorrelation is sufficient for the Gaussian model). Then, the following model is obtained:
\begin{equation}
\mathrm{vec}\begin{pmatrix}\alpha_{11} & \cdots & \alpha_{1p}\\\vdots & \ddots & \vdots\\ \alpha_{n1} & \cdots & \alpha_{np}\end{pmatrix}\sim\mathrm{Gaussian}\left(\begin{pmatrix}\mu_1 1_n\\\vdots\\\mu_p 1_n\end{pmatrix}, \begin{pmatrix}\Sigma_1 & \cdots & 0\\\vdots & \ddots & \vdots\\ 0 & \cdots & \Sigma_p\end{pmatrix}\right),
\end{equation}
where $1_n$ is the length-$n$ vector of ones and the $\Sigma_j\in\mathbb{R}^{n\times n}$ are stationary covariance matrices with $(\Sigma_j)_{ii'}=\sigma_j(s_i-s_{i'})$. Moreover, as before the gradient processes $\nabla \alpha_j$ are jointly normal with the above, and the same results as above hold for each of the $\alpha_j$, with independence between them.

When trying to classify a boundary as a wombling boundary when the spatial process is functional, a natural consideration is a functional extension of the normalised wombling measure: 
\begin{equation}
\bar{W}_Z^f(\mathcal{C}; t) := \frac{1}{|\mathcal{C}|}\int_\mathcal{C} n_\mathcal{C}^\top(s)\nabla Z(s; t)\mathrm{d}\ell(s).
\end{equation}
A one-dimensional summary of the flux of $\nabla Z$ across $\mathcal{C}$ over time is $||\bar{W}_Z^f(\mathcal{C})||^2$. Given an orthonormal basis expansion, $||\bar{W}_Z^f(\mathcal{C})||^2$ is readily computed as the $\ell^2$ squared norm of the sequence of loadings; given a truncated expansion has been decided, the truncated squared norm will serve as an appropriate approximation. Thus, one can jointly infer on each of the $\bar{W}_{\alpha_j}(\mathcal{C})$ for $j\in[p]$, as in the univariate methodology, and take this squared norm $\sum_{j=1}^p \bar{W}_{\alpha_j}(\mathcal{C})^2$ for the posterior sample of $||\bar{W}_Z^f(\mathcal{C})||^2$. Again, credible sets or standardised scores can be obtained for inference.

\section{Nonparametric Wombling}

Below is developed a novel nonparametric and fully functional approach to Wombling (although it may also be simplified to the multivariate or univariate case as well). 

\subsection{Framework}

Suppose the functional data $Y_i\in L^2(\mathcal{T})$ for some (not necessarily one-dimensional) domain $\mathcal{T}$ are generated according to the following model:
\begin{equation}
Y_i = Z(s_i) + \epsilon_i\quad \text{for}\quad i\in[n],\ s_i\in\mathcal{S}\subset\mathbb{R}^2
\end{equation}
where $Z:\mathcal{S}\longrightarrow L^2(\mathcal{T})$ is a mean square differentiable second-order stationary spatial functional field, i.e. it has moments
\begin{equation}
\mathbb{E}Z(s) = \mu\quad \text{and}\quad \mathbb{E}\left[(Z(s) - \mu)\otimes(Z(s') -\mu)\right] = c(s-s')\quad \text{for}\quad s,s'\in\mathcal{S}, \quad \text{with}\ c(0)=\lim_{\delta\rightarrow 0}c(\delta),
\end{equation}
where $\mu\in L^2(\mathcal{T})$, $c:\mathbb{R}^2\rightarrow \mathcal{B}(\mathcal{T})$ is continuous at $0$ (ie $Z$ is mean square differentiable), $\mathcal{B}(\mathcal{T})$ is the set of bilinear functions on $\mathcal{T}$ representing a valid kernel for a trace class positive definite operator on $L^2(\mathcal{T})$, $\otimes$ is now the outer product on the Hilbert space $L^2(\mathcal{T})$ (as opposed to $L^2(\mathcal{S})$); and $\epsilon_i$ is spatial homoscedastic white noise functional process, ie
\begin{equation}
\mathbb{E}\epsilon_i = 0\quad \text{and}\quad \mathbb{E}\left(\epsilon_i\otimes\epsilon_j\right) = \mathbb{I}(i=j)\nu\quad \text{for}\quad i,j\in[n],\quad \text{and}\quad (\epsilon_i)_{i\in[n]}\perp Z,
\end{equation}
where $\mathbb{I}$ is the indicator function, and $\nu\in\mathcal{B}(\mathcal{T})$. Note that this is a much more general framework than for the Bayesian methodology, which necessarily required a parametric family (i.e. Gaussianity) as well as model simplifications for finite-dimensional parametric inference.

Note that, much like in the Bayesian case, this framework and the following methodology can be simply extended to include a regression component, allowing mean-nonstationarity in the data; however this is not of immediate interest in the application, so it will be left out to simplify the notation and derivations.

The aim is now to predict $\nabla Z$, in particular the functional of $\nabla Z$ already defined as the normalised functional wombling measure:
\begin{equation}
\bar{W}_Z^f(\mathcal{C}; t) := \frac{1}{|\mathcal{C}|} \int_\mathcal{C} n_\mathcal{C}^\top (s) \nabla Z(s; t) \mathrm{d}\ell(s)\in L^2(\mathcal{T}),
\end{equation}
for some curve $\mathcal{C}$ in $\mathcal{S}$ with unit normal $n_\mathcal{C}(s)$ at point $s\in\mathcal{C}$; once again, $\ell$ represents the arc length measure, and $|\mathcal{C}|:=\int_\mathcal{C}\mathrm{d}\ell$ is the total arc length of curve $\mathcal{C}$.

\subsection{Methodology}

Since $\bar{W}_Z^f(\mathcal{C})$ is a linear functional of $Z$, it is appropriate to consider a predictor that is linear in the observed $Y_i$. In particular, sought is the best linear unbiased predictor (\textsc{blup}) of $\bar{W}^f_Z(\mathcal{C})$:
\begin{equation}
\text{minimise}\quad \mathbb{E}\left|\left|\sum_{i=1}^n \tilde{w}_iY_i - \bar{W}_Z^f(\mathcal{C})\right|\right|^2\quad \text{over}\quad \tilde{w}\in\mathbb{R}^n\quad \text{subject to}\quad \mathbb{E}\left(\sum_{i=1}^n \tilde{w}_i Y_i - \bar{W}^f_Z(\mathcal{C})\right) = 0,
\end{equation}
where $||\cdot||$ is the $L^2$ norm. Let $w\in\mathbb{R}^n$ denote the optimal such vector of weights.

By the stationarity of the process $Z$, $\mathbb{E}\nabla Z = 0$, and therefore the right-hand condition becomes $\mu=0$ or $\sum_{i=1}^n \tilde{w}_i=0$. If $\mu$ is known, the data can be pre-processed to centre the distribution of the $Y_i$, and the former assumption can be made; this is the case in the application since it is known that $\mu=\mathrm{id}$, and this improves the accuracy of the subsequent prediciton. However, for the sake of generality, for now it will not be assumed that $\mu$ is known, and the derivations will assume the restriction that $\sum_{i=1}^n\tilde{w}_i=0$. However, at the end the (simpler) form for $w$ in the case that $\mu$ is known and the data pre-processed will also be given.

The optimisation problem can now be solved by the method of Lagrange multipliers:
\begin{equation}
\text{minimise}\quad L(\tilde{w},\lambda):=\mathbb{E}\left|\left|\sum_{i=1}^n \tilde{w}_iY_i - \bar{W}_Z^f(\mathcal{C})\right|\right|^2 + \lambda\sum_{i=1}^n \tilde{w}_i.
\end{equation}
This provides the following conditions:
\begin{equation}
\left\{ \forall j\in[n],\quad 2\mathbb{E}\left\langle\sum_{i=1}^n w_iY_i - \bar{W}_Z^f(\mathcal{C}), Y_j\right\rangle + \lambda = 0\right\} \wedge \left\{ \sum_{i=1}^n w_i = 0 \right\}.
\end{equation}
This can be rearranged to provide the equivalent conditions in terms of the model parameters:
\begin{multline}
\left\{ \forall j\in[n],\quad 2\sum_{i=1}^n w_i \left( \tr c(s_i-s_j) + \mathbb{I}(i=j)\tr \nu\right) - \frac{2}{|\mathcal{C}|}\int_\mathcal{C} n_\mathcal{C}^\top(s) \nabla \tr c(s_j - s) \mathrm{d}\ell(s) + \lambda = 0 \right\} \wedge\\ \left\{ \sum_{i=1}^n w_i = 0 \right\}.
\end{multline}
Defining the matrix $C\in\mathbb{R}^{n\times n}$ with $C_{ij}=\tr c(s_i-s_j)$ and the vector $b\in\mathbb{R}^n$ with $b_i =\frac{1}{|\mathcal{C}|}\int_\mathcal{C} n_\mathcal{C}^\top(s) \nabla\tr c(s_i - s)\mathrm{d}\ell(s)$, this then says:
\begin{equation}
\left\{ (C + \tr \nu \mathrm{I}_n)w - b = -\frac{\lambda}{2}1_n \right\}\wedge\left\{ 1_n^\top w = 0\right\},
\end{equation}
where $\mathrm{I}_n$ is the $n\times n$ identity matrix and $1_n$ is the length-$n$ vector of ones. This linear system is then solved by:
\begin{equation}
w = (C + \tr\nu \mathrm{I}_n)^{-1} \left( b - \frac{1_n^\top (C+\tr\nu \mathrm{I}_n)^{-1} b}{1_n^\top (C+\tr\nu \mathrm{I}_n)^{-1} 1_n}1_n \right),
\end{equation}
providing the \textsc{blup} of the normalised functional wombling measure: $\sum_{i=1}^n w_iY_i$.

In the case $\mu$ is known, the above derivation simplifies greatly, providing $w = (C+\tr\nu \mathrm{I}_n)^{-1} b$, and the normalised functional wombling measure predictor is then $\sum_{i=1}^n w_i(Y_i - \mu)$.

The question remains however of how to estimate the model parameters; this will be done via a functional adaptation of classical geospatial statistical methodology. Note the following:
\begin{equation}
\frac{1}{2}\mathbb{E}||Y_i - Y_j||^2 = \tr (c(0) + \nu) - \tr (c(s_i - s_j) + \mathbb{I}(i=j) \nu).
\end{equation}
Thus, the data $\left(s_i - s_j, ||Y_i-Y_j||^2\right)$ for $i,j\in[n]$ and $i\neq j$ can be used as observations of this trace of the functional variogram $\gamma(s_i-s_j):=\frac{1}{2}\mathbb{E}||Y_i - Y_j||^2$. This should have a nugget component ($\tr\nu \mathbb{I}(\delta\neq 0)$) plus a covariance structure ($\gamma_0(\delta)$) implying mean square differentiability. For analytic computation of derivatives, for the simulations and application a linear combination of Gaussian models will be used, although other models may be used with numeric differentiation. The data (indexed by $(i,j)$) can be fitted to this model by various means, including via iterated weighted least squares with target weights $1/\gamma(s_i-s_j)^2$ as described by Cressie, producing a fitted estimate of the trace of the functional variogram of $\hat{\gamma}(\delta) = \widehat{\tr\nu}\mathbb{I}(\delta\neq 0) + \widehat{\gamma_0}(\delta)$. \cite{cressie1985fitting} Then, $\tr c$ is estimated as $\widehat{\tr c}(\delta) = \widehat{\gamma_0}(\infty) - \widehat{\gamma_0}(\delta)$, which can then be plugged into the above with $\widehat{\tr \nu}$ to produce estimates $\hat{w}$ of the optimal weights.

Once again, if $\mu$ is known, the above can be simplified. In particular, one can work with the covariogram rather than the variogram:
\begin{equation}
\mathbb{E} \langle Y_i-\mu, Y_j-\mu\rangle = \tr (c(s_i-s_j) + \mathbb{I}(i=j) \nu).
\end{equation}
Hence, the data $\left(s_i-s_j, \langle Y_i-\mu,Y_j-\mu\rangle\right)$ for $i,j\in[n]$ and $i\neq j$ can be used as observations of this trace of the functional covariogram, where a model akin to the above can be fitted and $\tr \nu$ and $\tr c$ can be estimated directly, which then feed into the estimates of the optimal weights $w$ via the simplified formula given above.

Now the normalised functional wombling measure $\bar{W}_Z^f(\mathcal{C})$ has been predicted, $||\bar{W}_Z^f(\mathcal{C})||^2$ can be computed as the one-dimensional summary of the functional wombling measure.

\subsection{Inference}

In order to account for potentially differing variances of the measures, Castillo--Páez et al.'s nonparametric bootstrap for spatial data has been developed for the functional framework, utilising an approximation that the functional and spatial covariance structures are separable. \cite{castillo2019nonparametric} In particular, given the estimated covariance matrix $M_{ij}:=\widehat{\tr c}(s_i-s_j)+\mathbb{I}(i=j)\widehat{\tr \nu}$, the Cholesky decomposition $M=LL^\top$ where $L$ is a lower-triangular matrix can be cheaply computed. Then, $Y_i^\pi:=\sum_{j=1}^n\sum_{k=1}^n L_{ij}\left(L^{-1}\right)_{\pi(j)k}Y_k$ for $i\in[n]$ and where $\pi$ is a random permutation of $[n]$ can be treated as a bootstrap sample. Then bootstrap estimates of $||\bar{W}_Z^f(\mathcal{C})||^2$ utilising the bootstrap samples can be computed and compared to the original estimate. (Note that these computations use the same estimated vector of weights $w$, and so this is computationally cheap to run.) Then, a pseudo--p value can be estimated by the rank of the original estimate among the bootstrap estimates, and this can be used to compare different curves' estimated measures.

\section{Simulations}

To compare the two developed functional wombling methodologies, simulations were run where the aim was to correctly identify a `true' boundary compared to a `false' boundary.

\subsection{The data and the boundaries}

For each simulation, first the observation locations $(x_i, y_i) = s_i\in [0,1]^2$ for $i\in[100]$ were generated unfiromly at random. Then, the functional data were generated according to the following Gaussian model:
\begin{equation}
F_i = \sum_{j=1}^{10} \alpha_{ij}\phi_j,\quad \begin{pmatrix}\alpha_{1j}\\\vdots\\\alpha_{100j}\end{pmatrix}\sim\mathrm{Gaussian}\left( \mu, \Sigma_j \right),
\end{equation}
where the $\phi_j$ are the Legendre polynomials transformed to the domain $[0,1]$. The form of the $\mu$ and the $\Sigma_j$ in the above are based on the form of the boundary to be identified. In particular, two types of boundaries are considered: a mean-like boundary and a covariance-like boundary. In both cases, to be compared is a null boudary on the segment from $(0.25,0.4)$ to $(0.25,0.6)$ and a real boundary on the segment from $(0.75,0.25)$ to $(0.75,0.75)$.

For the mean-like boundary, the covariance structure is of the form $(\Sigma_j)_{ii'} = \frac{\nu_j}{j^2} \mathrm{I}_{100} + \frac{\sigma_j}{j^2} \exp\left( - \frac{||s_i - s_{i'}||^2}{\rho_j^2} \right)$, where $\nu_j\sim\mathrm{Uniform}[0.05,0.15]$, $\sigma_j\sim\mathrm{Uniform}[0.8, 1.2]$ and $\rho_j\sim\mathrm{Uniform}[0.4,0.6]$ independently. The boundary comes from the mean structure, which takes the following form:
\begin{footnotesize}
\begin{equation}
\mu_i(f) = \begin{cases}
f \mathbb{I}(|x_i-\frac{1}{4}| < \frac{1}{4})\mathbb{I}(|y_i-\frac{1}{2}| < \frac{1}{10})\sign(x_i-\frac{1}{4})\left(\cos(2\pi(x_i-\frac{1}{4}))\right)^2\tanh\left(5\left(\cos(5\pi(y_i-\frac{1}{2}))\right)^2\right), & \text{left boundary};\\
f \mathbb{I}(|x_i-\frac{3}{4}| < \frac{1}{4})\mathbb{I}(|y_i-\frac{1}{2}| < \frac{1}{4})\sign(x_i-\frac{3}{4})\left(\cos(2\pi(x_i-\frac{3}{4}))\right)^2\tanh\left(5\left(\cos(2\pi(y_i-\frac{1}{2}))\right)^2\right), & \text{right boundary}.
\end{cases}
\end{equation}
\end{footnotesize}
This therefore forces an additional gradient across the real boundary of varying size depending on some boundary factor $f$, with no forcing on the null boundary, under which there is a stationary and isotropic Gaussian process.

For the covariance-like boundary, the mean structure is simply zero; all of the structure is in the covariances. These are of the familiar form $(\Sigma_j)_{ii'} = \frac{\nu_j}{j^2} \mathrm{I}_{100} + \frac{\sigma_j}{j^2} \exp\left( - \frac{||s_i^* - s_{i'}^*||^2}{\rho_j^2} \right)$, where $\nu_j\sim\mathrm{Uniform}[0.05,0.15]$, $\sigma_j\sim\mathrm{Uniform}[0.8, 1.2]$ and $\rho_j\sim\mathrm{Uniform}[0.4,0.6]$ independently. Here, the $s_i^*$ are the original locations $s_i$ but embedded now in $\mathbb{R}^3$, with an additional vertical component $z_i(f)$ of the same form as the $\mu_i(f)$ above. Thus, observations either side of the boundary have an inflated distance between them, reducing the correlation between the observed values. Note a Euclidean distance is retained and hence the covariance structure remains valid.

\subsection{The methodologies}

For the functional Bayesian approach, the generated functions were decomposed in a Fourier basis, and truncated at $p\in{1,3,10}$ components; independence of the loadings is then an approximation, as in any application. It was assumed that the loadings mean $\mu=0$ was known (the mean-like boundary being a deviation from this null), as it is extrinsically known in the Covid-19 wave phase variation application. The covariance structure for each component was assumed to be the sum of a nugget and a Gaussian term; the nuggets, semisills and ranges may be different for the different components. Largely non-informative priors defined on $(0,\infty)$ were imposed independently for each of the ${3,9,30}$ parameters, namely inverse gamma priors with shape parameter 2 and scale paramater 0.1 (so mean 0.1 and infinite variance).

For the nonparametric approach, again it was assumed that the functional mean $\mu=0$ was known, and for the covariance structure a nugget plus 1, 3 and 10 Gaussian terms were utilised, and thus 3, 7 and 21 total parameters, respectively. This covariance structure was fitted by nonlinear least squares, taking into account only the pairs of observations less than half the maximum pairwise distance apart.

\subsection{The results}

\begin{figure}
\centering
\includegraphics[width=0.7\textwidth]{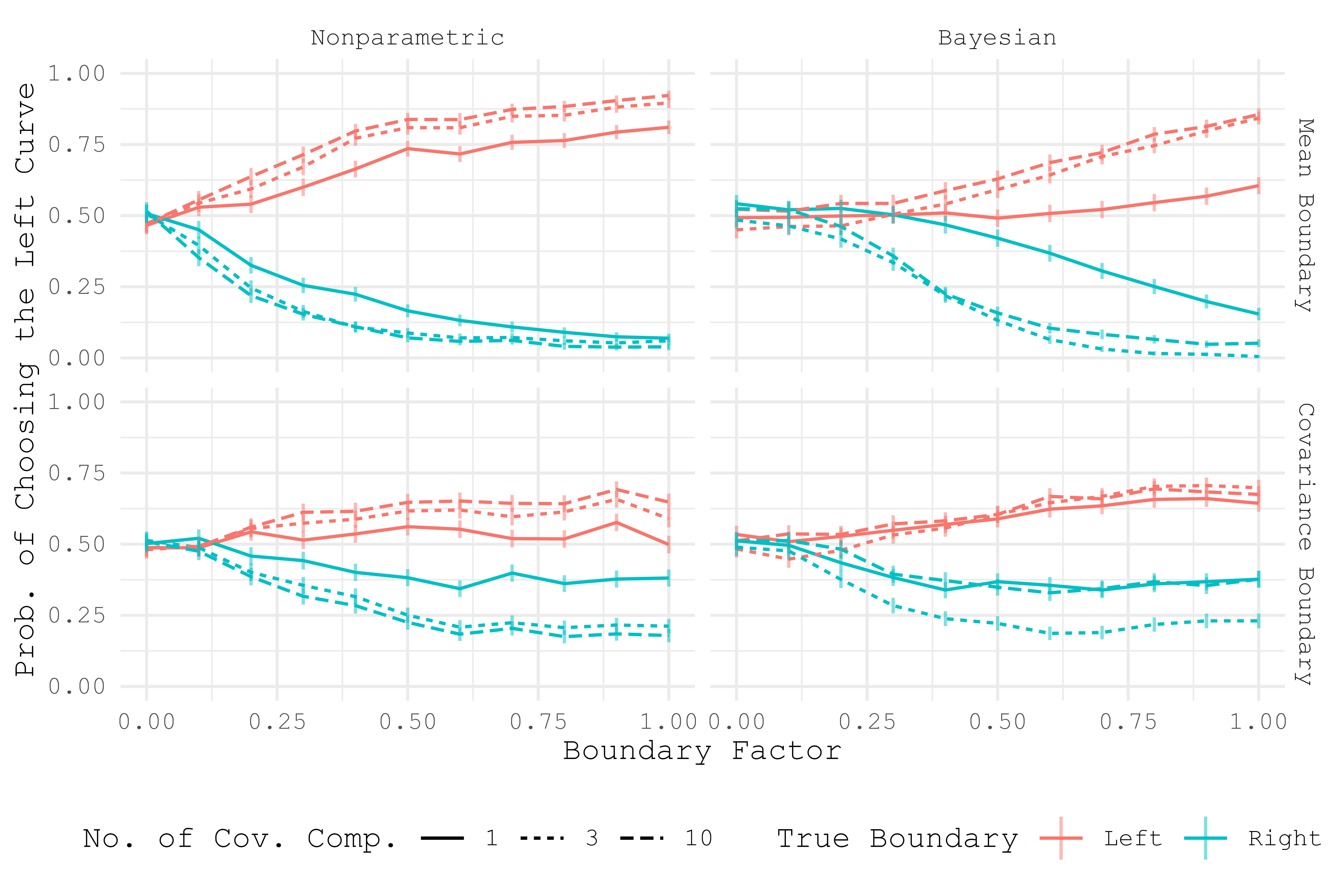}
\caption{Simulation Results: The probability of choosing the left line segment as the true boundary under the various methodologies for it being the true and null boundary. Provided also are the 95\% Wald confidence intervals for these probabilities. Choosing the boundary means: in the Bayesian case, that the standardised wombling measure (posterior mean divided by posterior standard deviation) was greater for that boundary; in the nonparametric case, that the bootstrap pseudo--p value is lesser for that boundary. The boundary factor represents the level of boundary introduced, represented by $f$ in the formulae above, hence for $f=0$ there is no boundary and so there shouldn't be a preference for either boundary.}\label{simulation_results}
\end{figure}

The results of the simulations are given in Figure \ref{simulation_results}, from 1024 simulations. It can be seen, especially for when the boundary arises from the mean structure of the data generation, the nonparametric methodology is much more likely to choose the correct boundary as exhibiting the greatest wombling measure at much lower levels of boundary factor $f$. Moreover, the particular implementation in this case saw the nonparametric methodology an order of magnitude faster to run than the Bayesian methodology. 

\section{Covid-19 Wave Phase Variation}

The nonparametric methodology above was applied to the phase variation of the first wave of Covid-19 in the United Kingdom. In particular, this is the inferred data $H_i^{-1}$ from the model $X_i = \Xi_i \mu\circ H_i^{-1}$, where $X_i$ is the trajectory of SARS-CoV-2 incidence in calendar time in local authority $i$, $\mu$ is some central wave shape with $\Xi_i$ a multiplicative factor allowing rank-one variation in amplitude, and $H_i$ are the registration functions (increasing bijections on $\mathcal{T}$) transforming time from a latent global clock in which features are aligned into local calendar time. The $H_i$ therefore represent phase variation of the incidence trajectories, encoding how early/late or sharp/flat the waves were when aggregated in each local authority. These $H_i$ have been extracted from daily data of positive SARS-CoV-2 test results from March to June 2020 disaggregated by 380 local authorities in the UK, as in the methodology of Barratt and Aston. \cite{barratt2024exploring} As their paper demonstrates, there are clear qualitative and quantitative differences in this phase variation of the Covid waves between local authorities in London and outside of London, suggesting a wombling analysis of this boundary with respect to the estimated phase variation functions $H_i$. It should be noted that $\mathbb{E}H_i=\mathrm{id}$ by assumption, for the sake of identifiability of phase and amplitude variation in the model, and so it should be taken that the mean is known.

\begin{figure}
\centering
\includegraphics[width=0.7\textwidth]{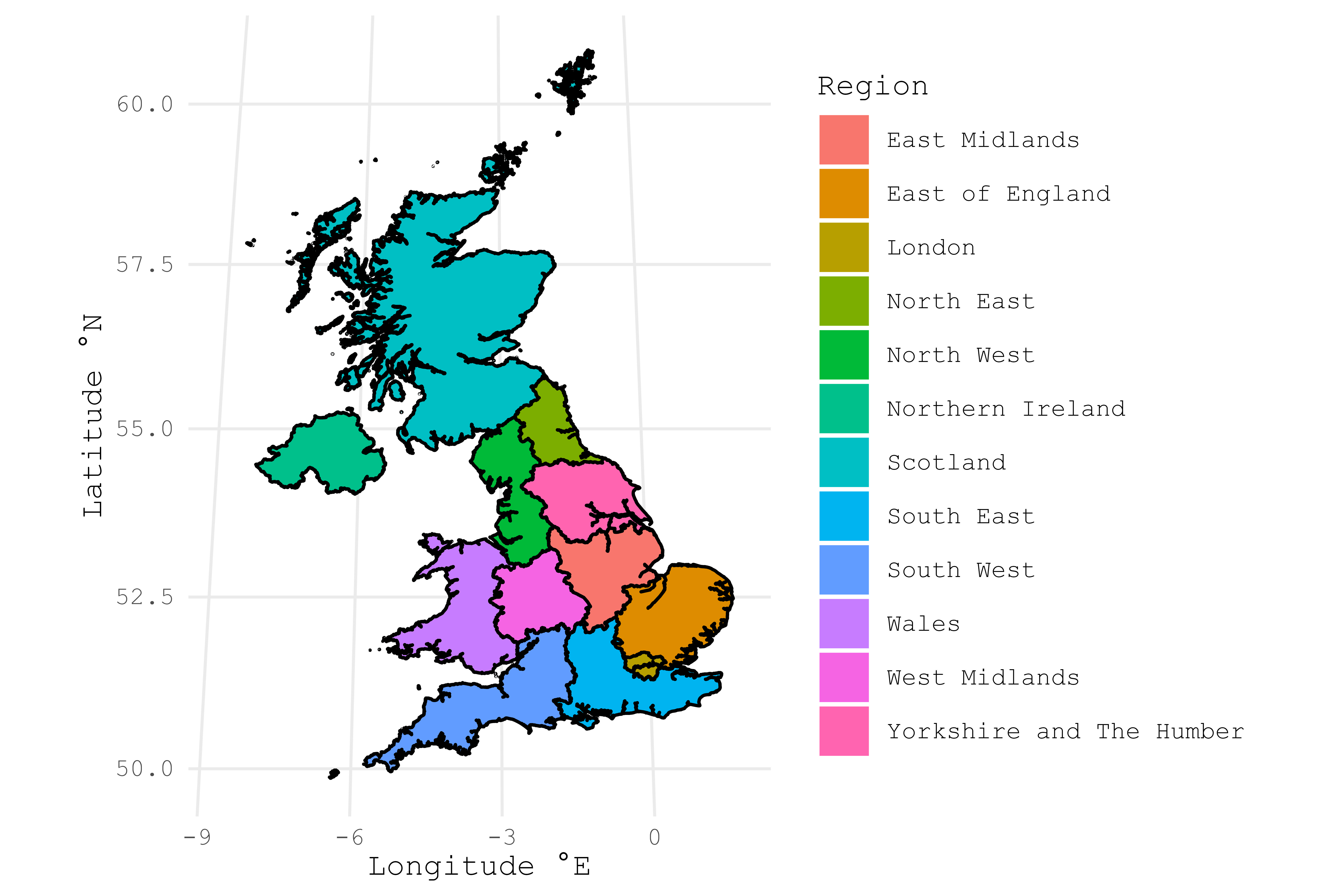}
\caption{Regions of the United Kingdom.}\label{regions}
\end{figure}

\begin{table}
\centering
\begin{tabular}{r|lll}
Boundary & Measure & p Value & $\pm$\\\hline
East--Ldn &1.0235e-05 &  0.035& 0.0113908\\
East--SE &3.9973e-07  & 0.927 &0.0161234\\
EMid--East &7.8533e-07 &  0.684& 0.0288156\\
EMid--SE &1.6035e-06  & 0.412 &0.0305066\\
EMid--WMid &1.2100e-06 &  0.486& 0.0309782\\
EMid--YatH &7.5227e-07 &  0.540& 0.0308910\\
Ldn--SE &9.7177e-06   &0.021 & 0.0088870\\
NE--NW &6.6930e-07   &0.441& 0.0307738\\
NE--Scot &1.1282e-06  & 0.291 &0.0281531\\
NE--YatH &5.4124e-08  & 1.000& 0.0000000\\
NW--EMid &5.8066e-07 &  0.978& 0.0090915\\
NW--Scot &2.6942e-06  & 0.137& 0.0213119\\
NW--Wales &6.6164e-07 &  0.861& 0.0214420\\
NW--WMid &3.2240e-07 &  0.988& 0.0067488\\
NW--YatH &1.0454e-06  & 0.514 &0.0309782\\
SE--SW &3.6433e-07  & 0.777 &0.0258000\\
SW--Wales &3.5334e-06 &  0.611 &0.0302170\\
WMid--SE &1.6242e-06  & 0.683 &0.0288401\\
WMid--SW &1.1368e-07 &  0.990 &0.0061670\\
WMid--Wales& 6.7268e-07&   0.319 &0.0288885\\\hline
\end{tabular}
\caption{Results of Covid Application Computations: The estimated wombling measures along with the associated pseudo--p values for each land region--region boundary in the United Kingdom. Also provided are Wald confidence intervals for the p values based on the bootstrap sampling size.}
\label{covid_results}
\end{table}

Figure \ref{regions} provides a map of the regions of the UK for reference, which are subdivided into local authorities. For each land boundary between regions, the functional wombling measure with respect to the $H_i$ treated as observed at the centroids of the local authorities was predicted, and bootstrap samples made to convert these measures into pseudo--p values. The results of these computations are provided in Table \ref{covid_results}. These results demonstrate that indeed the London boundaries are significant in comparison to other regional boundaries in the UK, as well as having the greatest predicted wombling measure. Also of note is the impact of the bootstrapping, which sees certain boundaries such as the South West--Wales boundary which have relatively large predicted wombling measures lose this exceptionalism in their p value, and vice versa. In this case this can be understood since the South West--Wales boundary is exceptionally short, and so the standard error in estimating the wombling measure is larger than for longer boundaries.

This analysis demonstrates that the shape of the incidence trajectories indeed differs substantially within and outside the London boundary, which may motivate separate analysis for London epidemiology compared to the rest of the UK. Moreover, in public policy terms, it justifies the claims of the time---largely based off anecdotal or scarce evidence---that London potentially should have been treated differently in terms of intervention: in particular, stricter lockdown measures, or inter-region travel restrictions. In future, this understanding of epidemiological behaviour in London may reinforce this evidence.
\begin{quotation}
It is now clear that the peak of the epidemic is coming faster in some parts of the country than in others. It looks as though London is now a few weeks ahead... So to relieve the pressure on the London health system and to slow the spread in London, it is important that Londoners now pay special attention to what we are saying about avoiding non-essential contact and to take particularly seriously the advice about working from home and avoiding confined spaces such as pubs and restaurants.
\flushright ---Prime Minister Boris Johnson, March 2020
\end{quotation}

\section{Discussion}

This paper has seen the development of a novel methodology for wombling, namely a nonparametric approach that can be applied in the case of functional data. When compared to a functional extension of the established Bayesian methodology of Banerjee and Gelfand, several comparisons can be drawn. First, whilst Banerjee and Gelfand's methodology has proven success and popularity, throughout it will become clear limitations of this approach in a more general setting: in particular, there is a requirement of a Gaussianity assumption along with simplifying assumptions on the covariance structure to keep the computations feasible, although (particularly in the functional case) they remain costly both in the running of the \textsc{mcmc} posterior estimation and the subsequent linear algebra and quadrature required for each sample from the \textsc{mcmc}. Meanwhile, the nonparametric approach far outperforms the Bayesian approach computationally, with requiring minimal modelling assumptions.

The nonparametric approach does lose the convenient probabilistic inference of the Bayesian methodology, which may be useful for gradient estimation where confidence/credible intervals may be useful; however, for wombling in practice comparison to other curves is what is required, which can be achieved in this methodology, as it can in the Bayesian methodology, by standardisation of the estimates by a measure of the scale of their variance. In simulation analysis the novel methodology does moreover prove superior at identifying boundaries than the Bayesian methodology. Finally, note that this approach can be applied in the univariate case as well, although many of the advantages compared to the Bayesian methodology are lessened; in the (finitely) multivariate case, depending on the application and what may be extrinsically justified, either methodology may be more attractive.

The application provides an example of the ability of the nonparametric functional methodology to draw interpretable conclusions in a real-world data setting. In particular, it successfully identifies London as an anomalous region when it comes to the shapes of Covid-19 incidence trajectories, which reflect both the analysis of the original Barratt and Aston paper, as well as comments made at the time based off limited data. This analysis is therefore relevant to retrospective discussion of the Covid-19 epidemic in the UK with respect to public policy decision making, as well as an indicator that London perhaps should be considered separately in future epidemics.

One aspect that could be developed is the following fact. Note that if the flux across $\mathcal{C}$ reverses direction for some subset of the functional domain $\mathcal{T}$, but remains of similar magintude, this measure is unchanged, unlike along the curve $\mathcal{C}$ when the flux has to be in the same direction. In the latter case, it appears more sensible to consider the curve $\mathcal{C}$ two wombling boundaries, separated by further wombling boundaries; whereas, in the former case, it means a truly functional distance on $\mathcal{T}$ across the boundary is considered rather than some average of the difference at each point in $\mathcal{T}$. However, depending on the application, this may not be the case, and indeed various other forms of wombling measure, or how a one-dimensional summary may be obtained from a functional wombling measure, may be considered. This does cause difficulty in both methodologies for the inference on this alternative wombling measure, which should be considered.

\bibliographystyle{plain}
\bibliography{ref}

\begin{thebibliography}{1}

\bibitem{banerjee2006bayesian}
Sudipto Banerjee and Alan~E Gelfand.
\newblock Bayesian wombling: Curvilinear gradient assessment under spatial
  process models.
\newblock {\em Journal of the American Statistical Association},
  101(476):1487--1501, 2006.

\bibitem{barratt2024exploring}
Luke~A Barratt and John~AD Aston.
\newblock Exploring covid-19 spatiotemporal dynamics: Non-euclidean spatially
  aware functional registration.
\newblock {\em arXiv preprint arXiv:2407.17132}, 2024.

\bibitem{castillo2019nonparametric}
Sergio Castillo-P{\'a}ez, Rub{\'e}n Fern{\'a}ndez-Casal, and Pilar
  Garc{\'\i}a-Soid{\'a}n.
\newblock A nonparametric bootstrap method for spatial data.
\newblock {\em Computational statistics \& data analysis}, 137:1--15, 2019.

\bibitem{cressie1985fitting}
Noel Cressie.
\newblock Fitting variogram models by weighted least squares.
\newblock {\em Journal of the international Association for mathematical
  Geology}, 17:563--586, 1985.

\bibitem{womble1951differential}
William~H Womble.
\newblock Differential systematics.
\newblock {\em Science}, 114(2961):315--322, 1951.

\end{thebibliography}

\end{document}